\documentclass{amsart}

\def\lam#1{\lambda_{#1}}
\def\mshift#1{\{#1\}}
\def\C{{\mathbb C}}
\def\t{\textbf{t}}
\def\I{{\mathbb I}}
\def\calL{\mathcal{L}}
\def\S{{\bf S}}
\def\Z{{\mathbb Z}}
\newtheorem{theorem}{Theorem}[section]
\newtheorem{corollary}[theorem]{Corollary}
\newtheorem{lemma}[theorem]{Lemma}

\def\mybibcite[#1]{\cite{#1}}
\def\note{\par\medskip\noindent\textbf{Note:} }
 \makeatletter
\def\@oddhead{\underline{\hbox to \textwidth{On KP Generators and the Geometry of the HBDE \hfill Gekhtman/Kasman}}}
\let\@evenhead\@oddhead \def\@oddfoot{$\overline{\hbox to
\textwidth{\hfill \bf\thepage\hfill}}$} \let\@evenfoot\@oddfoot
\advance\footskip.5in \makeatother

\begin{document}
\title{On KP Generators and the Geometry of the HBDE}
\author{Michael Gekhtman}
\address{Department of Mathematics / University of Notre Dame / South
  Bend, IN / USA}

\author{Alex Kasman}
\address{Department of Mathematics / College of Charleston /
  Charleston, SC / USA}

\begin{abstract}
Sato theory provides a correspondence between solutions to the KP
hierarchy and points in an infinite dimensional Grassmannian.  In this
correspondence, flows generated infinitesimally by powers of the
``shift'' operator give time dependence to the first coordinate of an
arbitrarily selected point, making it a tau-function.  These
tau-functions satisfy a number of integrable equations, including the
Hirota Bilinear Difference Equation (HBDE).  Here, we
rederive the HBDE as a statement about linear maps between
Grassmannians.  In addition to illustrating the fundamental nature of
this equation in the standard theory, we make use of this geometric
interpretation of the HBDE to answer the question of what
\textit{other} infinitesimal generators could be used for similarly
creating tau-functions.  The answer to this question involves a ``rank one condition'', tying this investigation to the existing results on integrable systems involving such conditions and providing an interpretation for their significance in terms of the relationship between the HBDE and the geometry of Grassmannians.
\end{abstract}
\maketitle
\section{Introduction}

It was the seminal work of Sato \cite{Sato} which related the
geometry of the Grassmannian to the solution of soliton equations.
That relationship is analogous to the relationship of the functions
sine and cosine and the geometry of the unit circle in the
plane. These trigonometric functions, of course, arise as the
dependence of the $x$ and $y$-coordinates on the time parameter of a uniform
flow around the circle.  In the case of Sato theory, it is the
tau-functions of the KP hierarchy which arise as the dependence of the
``first" Pl\"ucker coordinate upon the time variables
$\t=(t_1,t_2,t_3,\ldots)$ where the flow corresponding to the variable
$t_i$ is generated infinitesimally by the operator which takes the
basis element $e_j$ of the underlying vector space to $e_{j+i}$
\cite{Sato,SegalWilson,Willox}.

The remainder of this introduction will briefly review this
construction and motivate the following question: What \textit{other}
choice of infinitesimal generator could have been made that similarly generate KP tau-functions?
 In other
words, we are looking for other flows, in both finite
and infinite dimensional Grassmannians, which have this
property of creating tau-functions through the projection onto the
first coordinate.

Our approach to this question will be algebro-geometric in nature,
rather than analytic.  In Section~\ref{sec:HBDE} we
will reinterpret the Hirota Bilinear Difference Equation, which
characterizes KP tau-functions, as a linear map between Grassmann
cones with certain geometric properties.  It will be precisely the
existence of such a map that characterizes the alternate KP generators.

The main result appears in Section~\ref{sec:KPgen} where we identify those operators
$S$ that can serve as 
generators of the KP flow in a Grassmannian.  As it
turns out, this property is characterized only by a restriction on the
rank of one block of the operator.
This result is applied and discussed in Sections~\ref{sec:Appl} and
\ref{sec:Conclusions}, with special emphasis on its relationship to
the rank one conditions that have appeared elsewhere in the literature
on integrable systems.

\subsection{The KP Hierarchy}\label{sec:KP}

The KP hierarchy is usually considered as an infinite set of
compatible dynamical systems on the space of monic pseudo-differential
operators of order one.
A \textit{solution} of the KP hierarchy is any
pseudo-differential operator of the form
\begin{equation}
\calL=\partial + w_1(\t)\partial^{-1}+
w_2(\t)\partial^{-2}+\cdots\qquad \t=(t_1,t_2,t_3,\ldots)
\label{eqn:Laxopr}
\end{equation}
satisfying the
evolution equations
\begin{equation}
\frac{\partial}{\partial t_i}\calL=[\calL,(\calL^i)_+]\qquad i=1,2,3,\ldots\label{eqn:KPhier}
\end{equation}
where the ``$+$'' subscript indicates projection onto the differential
operators by simply eliminating all negative powers of $\partial$, and
$[A,B]=A\circ B- B\circ A$.  

Remarkably, there exists a convenient way to encode all information
about the KP solution $\calL$ in a single function $\tau(\t)$
satisfying certain bilinear differential equations.  Specifically,
each of the coefficients $w_i$ of $\calL$ can be written as a certain
rational function of $\tau(t_1,t_2,\ldots)$ and its derivatives
\cite{SegalWilson}.  Alternatively, one can construct $\calL$
from $\tau$ by letting $W$ be the pseudo-differential operator
$$
W=\frac{1}{\tau}
\tau(t_1-\partial^{-1},t_2-\frac12\partial^{-2},\ldots)
$$ and then $\calL:=W\circ\partial\circ W^{-1}$ is a solution to the
KP hierarchy \cite{AdlervanMoerbeke}.  Every solution to the KP
hierarchy can be written this way in terms of a tau-function, though
the choice of tau-function is not unique.  For example, note that one
may always multiply $W$ on the right by any constant coefficient
series $1+O(\partial^{-1})$ without affecting the corresponding
solution.  

If $\calL$ is a solution to the KP hierarchy then the function
$$u(x,y,t)=-2\frac{\partial}{\partial
x}w_1(x,y,t,\ldots)=2\frac{\partial^2}{\partial x^2}\log \tau$$ is a
solution of the KP equation which is used to model ocean waves.
Moreover, many of the other equations that show up as particular
reductions of the KP hierarchy have also been previously studied as
physically relevant wave equations.  The KP hierarchy also arises in
theories of quantum gravity \cite{vM}, the
probability distributions of the eigenvalues of random matrices
\cite{newAvM,Tracy}, and has applications to questions of classical
differential geometry \cite{doliwa}.

Certainly one of the most significant observations
regarding these equations, which is a consequence of the form
\eqref{eqn:KPhier}, is that all of these equations are completely
integrable.  Among the many ways to solve the equations of the KP
hierarchy are several with connections to the algebraic geometry of
``spectral curves'' \cite{Itsbook,Kr,Mul,PW,Pr,Shiota,SegalWilson}.
However,
more relevant to the subject of this note is the observation of
M. Sato that the geometry of an \textit{infinite dimensional
Grassmannian} underlies the solutions to the KP hierarchy
\cite{Sato}.

\subsection{Finite and Infinite Dimensional Grassmann Cones}

 Let $k$ and $n$ be two positive integers with $k<n$. For later
convenience, we will choose a non-standard notation for the basis of
$\C^n$, denoting it by
$$
 \C^n=\langle
e_{k-n},e_{k-n+1},\ldots,e_{-1},e_0,e_1,\ldots,e_{k-1}\rangle.$$
Then, for instance, an arbitrary element of ``wedge space''
$\bigwedge^k\C^n$ can be written in the form
$$
\omega=\sum_{I\in\I_{k,n}} \pi_I e_I
$$
where $\pi_I\in\C$ are coefficients, 
$\I_{k,n}$ denotes the
set
$$
\I_{k,n}=\left\{I=(i_0,i_1,\ldots,i_{k-1})|k-n\leq
  i_0<i_1<i_2<\cdots<i_{k-1}\leq k-1\right\}
$$
and $e_I=e_{i_0}\wedge e_{i_1}\wedge \cdots e_{i_{k-1}}$.

A linear operator $M:\C^n\to\C^n$ naturally
extends to an operator $\hat M: \bigwedge^k\to  \bigwedge^k$ where we consider
the action to be applied to each term of the wedge product
$$
\hat M e_I=M(e_{i_1})\wedge M(e_{i_2})\wedge \cdots \wedge M(e_{i_k})
$$
and extend it linearly across sums.

We denote by $\Gamma^{k,n}\subset \bigwedge^k \C^n$ the set of
decomposable $k$-wedges in the exterior algebra of $\C^n$:
$$
\Gamma^{k,n}=\left\{v_1\wedge v_2\wedge \cdots\wedge
v_k|v_i\in\C^n\right\}.
$$
This
\textit{Grassmann cone} is in
fact an affine variety in the ${n \choose k}$-dimensional vector space
$\bigwedge^k\C^n$ because $\omega$ is in $\Gamma^{k,n}$
if and only if the coefficients $\pi_I$ satisfy a collection of
quadratic polynomial relations known as the Pl\"ucker relations
\cite{HodgePedoe}.  
Specifically, we consider the coefficients $\pi_I$ to be
skew-symmetric in the ordering of their subindices and select any two
sets $I$ and $J$ of integers between $k-n$ and $n$
of cardinality $k-1$ and $k+1$, respectively:
$$
k-n\leq i_1<i_2<\cdots< i_{k-1} \leq n
$$
$$
k-n\leq j_1<j_2<\cdots<\leq j_{k+1} \leq n.
$$
It follows that $\omega$ is decomposable if and only if 
\begin{equation}
\sum_{l=1}^{l+1} (-1)^l \pi_{i_1,i_2,\ldots,i_{k-1},j_l}
\pi_{j_1,j_2,\ldots,j_{l-1},j_{l+1},\ldots,j_{k+1}}=0
\label{plucker}
\end{equation}
for all
such selections of subsets $I$ and $J$.

In general, therefore, the Grassmann cone $\Gamma^{k,n}$ is defined by a collection of
quadratic equations involving up to $k+1$ terms.  In the
special case
$k=2$ and $n=4$, only a single 3-term relation is required.
Specifically, $\omega\in\bigwedge^2\C^4$ is decomposable if and
only if the coefficients satisfy the equation
\begin{equation}
\pi_{-2,-1}\pi_{0,1}-\pi_{-2,0}\pi_{-1,1}+\pi_{-2,1}\pi_{-1,0}=0\label{gr24}.
\end{equation}
Later we will demonstrate a method through which 
the one relation \eqref{gr24} is sufficient to characterize the general case
(cf.\ Section~\ref{sec:alggeom}).

It is natural to associate  a $k$-dimensional subspace
$W_{\omega}\subset\C^n$ to a non-zero element
$\omega\in\Gamma^{k,n}$.  If
$\omega=v_1\wedge\cdots\wedge v_k$ then the $v_i$ are linearly
independent and we associate to $\omega$ the subspace $W_{\omega}$
which they span.  In fact, since $W_{\omega}=W_{\omega'}$
if $\omega$ and $\omega'$ are scalar multiples, it is more common to
consider the Grassmannian $Gr(k,n)={\mathbb P}\Gamma^{k,n}$ as a
projective variety whose points are in one-to-one correspondence with
$k$-dimensional subspaces.  This association of points in ${\mathbb
  P}\Gamma^{k,n}$ to $k$-dimensional subspaces is the
\textit{Pl\"ucker embedding} of the Grassmannian in projective space.  However, due to our interest in linear maps between these spaces -- and our desire to avoid having to deal with the complications of viewing them as rational maps between the corresponding projective spaces -- we choose to work with the cones instead.

Next, we briefly introduce the infinite dimensional
Grassmannian of Sato theory and the notation which will be most useful
in proving our main results.  Additional information can be found by
consulting \cite{KacPeterson,Miwa,Sato,SegalWilson}.

We formally consider the infinite dimensional Hilbert space $H$ over
$\C$ with basis $\{e_i|i\in \Z\}$.  It has the decomposition
\begin{equation}
H=H_-\oplus H_+
\label{H-+H+}
\end{equation}
where $H_-$ is spanned by $\{e_i|i<0\}$ and $H_+$ has the basis $\{e_i|i\geq0\}$.

The wedge space $\bigwedge$ has
the basis
$
e_I=e_{i_0}\wedge e_{i_1}\wedge \cdots$ where the (now infinite) multi-index $I=(i_0,i_1,i_2,\ldots)$ is selected from the set $\I$ 
whose elements are characterized by the properties $i_j<i_{j+1}$ and $i_j=j$ for $j$ sufficiently large.  (In other words, $I\in\I$ can be constructed from the ``ground state" $I_0=(0,1,2,3,4,\ldots)$ by selecting a finite number of its elements and replacing them with distinct, negative integers.)
A general element of $\bigwedge$ then is of the form
$$
\omega=\sum_{I\in\I}\pi_I e_I.
$$

Since the multi-indices are of this form, it is notationally convenient to write only the first $m$ elements
of an element of $I\in\I$ if it is true that $i_j=j$ for all $j\geq m$.
For instance, we utilize the abbreviations
$$
\pi_{-2,-1}=\pi_{-2,-1,2,3,4,5,\cdots}\qquad \textup{and}\qquad e_{-2,0,1}=e_{-2}\wedge
e_0\wedge e_1\wedge e_3\wedge e_4\wedge\cdots$$
and $e_{0,1}=e_0\wedge e_1\wedge e_2\wedge
e_3\wedge\cdots$.  Moreover, using this same abbreviation we are able
to view the finite set $\I_{k,n}$ introduced earlier as being a subset of the infinite $\I$:
$$
\I_{k,n}=\{I\in\I: -k<i_0,\ i_j=j\hbox{ for }j>n-1\}.
$$
In this way, arbitrary finite dimensional Grassmann cones can be seen as being embedded in the infinite dimensional one in the form of points with only finitely many non-zero Pl\"ucker coordinates.  Consequently, although we may not always emphasize this fact, the results we determine for $\bigwedge$ can all be stated in the finite dimensional case as well through this correspondence.

The Sato Grassmann cone $\Gamma \subset\bigwedge$ is precisely the set of
those elements which can be written as 
$$
\omega=v_1\wedge v_2\wedge v_3\wedge\cdots \qquad v_i\in H.
$$ It can also be characterized by Pl\"ucker relations since
$\omega\in \Gamma$ if and only if for every choice of $k$ and $n$, the
${n\choose k}$ Pl\"ucker coordinates $\pi_I$ for $I\in \I_{k,n}$
satisfy the relations \eqref{plucker} for $\Gamma^{k,n}$.  In order that the operations we are to utilize be well defined, we make the assumption that if $\omega$ is represented in this form, the vectors $\{v_i\}$ are chosen so that $v_i=e_i+\sum_{j=i+1}^{\infty} c_j e_j$ for $i$ chosen to be sufficiently large.

As in the finite dimensional case, the Grassmannian $Gr={\mathbb P} \Gamma$ has
an interpretation of being the set of subspaces of $H$ meeting certain
criteria.  However, rather than being identified by their dimension,
one can say that they are the subspaces for which the kernel and
co-kernel of a certain projection map are finite dimensional and for
which the index of that map is zero \cite{Sato,SegalWilson}.
Again, the subspace corresponding to $v_0\wedge v_1\wedge v_2\wedge
\cdots\in\Gamma$ is the subspace spanned by the basis $\{v_i\}$.

\note Those uncomfortable with the formal approach to this infinite
dimensional object may choose to assume further restrictions on these
definitions as specified in \cite{SegalWilson} where an
analytic approach is used to ensure that all objects are well defined
and that all infinite sums converge.  Alternatively, one may consider
the case that $\pi_I=0$ for $I\not\in \I_{k,n}$ in which case this
reduces to the finite dimensional situation in which there are no
questions of convergence.

\subsection{The Shift Operator and Tau-Functions}\label{sec:satoconstr}

The linear ``shift'' operator $\S:H\to H$ is defined by the property that
$\S e_i=e_{i+1}$. (Written as a matrix, it would have ones
on the sub-diagonal and zeros everywhere else.) 
The linear map 
\begin{equation}
E(\t)=\textup{exp}\sum_{i=1}^{\infty} t_i \S^i:H\to H,
\label{eqn:exp}
\end{equation}
 induces a 
map $\hat E(\t)$ on $\bigwedge$ for any fixed values of the parameters $\t=(t_1,t_2,\ldots)$.
We use $\hat E(\t)$ to introduce ``time dependence'' to each point $\omega\in\bigwedge$
\begin{equation}
\tilde\omega(t)=\hat E(\t) \omega=\sum_{I\in\I}\tilde\pi_I(\t) e_I.
\label{eqn:tildeomega}
\end{equation}

The main object of Sato's theory \cite{Sato} is the function
$\tau_{\omega}(\t)$ associated to any point $\omega\in\bigwedge$ and
is defined as the first Pl\"ucker coordinate of the time-dependent
point $\tilde\omega(\t)$ (cf.\ \eqref{eqn:tildeomega}):
\begin{equation}
\tau_{\omega}(\t)=\tilde\pi_{0,1}(\t).\label{eqn:taudef2}
\end{equation}
There is very little that one can say about 
$\tau_{\omega}(\t)$ in general.  In fact, since it can also be
described as an infinite sum of Schur polynomials with the original
coefficients $\pi_I$ of $\omega$ as coefficients
\cite{Sato,SegalWilson}, one can select $\omega\in\bigwedge$ so
that $\tau_{\omega}(\t)$ is any formal series in the variables $t_i$.

The main result of Sato theory is that $\tau_{\omega}(\t)$
is a KP tau-function precisely when $\omega\in\Gamma$.  In fact, a
function $\tau(\t)$ is a tau-function of the KP Hierarchy if and only if
$\tau(\t)=\tau_{\omega}(\t)$ for some $\omega\in\Gamma$ \cite{Sato}.

\note By virtue of the fact that we have chosen to work with Grassmann cones rather than projective Grassmannians, our correspondence between points and tau-functions necessarily involves the constant function $\tau_{0}(\t)\equiv0$.  The usual definition of  ``KP tau-function'' specifically excludes this function, but here we will adopt the convention of referring to this function as a KP tau-function even though it does not correspond in the usual way to a Lax operator $\calL$.

\subsection{Alternative KP Generators}

The main question which we seek to address in this paper is the
following: \textit{With what operator could you replace $\S$ in \eqref{eqn:exp} so that $\tau_{\omega}$ \eqref{eqn:taudef2} would still be a tau-function for any $\omega\in\Gamma$?}  

There is a sense in which this question seems uninteresting.  After all,
since Sato theory characterizes the totality of solutions of the KP
hierarchy using only the shift operator $\S$, it may not be clear why
one would be interested in other choices.  We therefore motivate the
question with the following list:
\begin{itemize}
\item   It is only by answering the question posed that we can
recognize which of the many properties that characterize the operator $\S$ are responsible for its
role in generating KP tau-functions.  For instance, it has the properties that it is a strictly lower triangular operator with respect to the basis $\{e_i\}$.  Additionally, it has the property that for $v\in H_-$, $\S v \in H_-\oplus \C e_0$.  It is not at first clear which, if any, of these properties is related to its role in generating KP flows.

\item Although all solutions of the KP hierarchy can be generated
using the operator $\S$ and some point $\omega\in\Gamma$ through Sato's
construction, it is possible that solutions which are difficult to
write or compute explicitly in that format can be derived in a simpler
way using an alternative choice of generator for the flows.  For
instance, the simplest points in $\Gamma$ are those having only
finitely many non-zero Pl\"ucker coordinates.  (Equivalently, one may
consider the case in which a finite dimensional Grassmannian is used
in place of the infinite-dimensional Sato Grassmannian.)  Using powers
of the shift operator $\S$ to generate the KP flows, these correspond
to tau-functions which are polynomials, depending only only a
finite number of the variables $\{t_i\}$ \cite{SegalWilson}.
However, as we will show, using an alternative generator one gets a wider variety of interesting KP tau-functions using flows
on \textit{finite dimensional} Grassmannians.

\item Finally, the answer to the question posed might provide an
understanding of other phenomena in integrable systems which were not
previously considered in the context of choice of KP generator in the Grassmannian at all.
In particular, we suggestively point out that ``rank one conditions''
(the requirement that a certain matrix have rank of at most one) have
arisen in the study of both finite and infinite dimensional integrable
systems in a number of apparently unrelated contexts.  We will argue
that these are related and actually represent an unrecognized instance
of the sort of alternative KP generator we we investigate here.
  
\end{itemize}

\section{The Geometry of the Hirota Bilinear Difference Equation}\label{sec:HBDE}

Although differential equations satisfied by KP tau-functions have certainly attracted the most attention, tau-functions are also known to satisfy \textit{difference} equations.  For instance, a tau-function $\tau(\t)$ necessarily satisfies the \textit{Hirota Bilinear Difference Equation} (HBDE) \cite{Miwa,Sato}
\begin{equation}\begin{matrix}
0&=&(\lam2-\lam1)(\lam4-\lam3)\tau(\t+\mshift{\lam1}+\mshift{\lam2})\tau(\t+\mshift{\lam3}+\mshift{\lam4})\cr
&&-(\lam3-\lam1)(\lam4-\lam2)\tau(\t+\mshift{\lam1}+\mshift{\lam3})\tau(\t+\mshift{\lam2}+\mshift{\lam4})\cr
&&+(\lam4-\lam1)(\lam3-\lam2)\tau(\t+\mshift{\lam1}+\mshift{\lam4})\tau(\t+\mshift{\lam2}+\mshift{\lam3})\cr
\end{matrix}
\label{HBDE}
\end{equation}
where the ``Miwa shift'' of the time variables $\t=(t_1,t_2,\ldots)$
is defined as\footnote{This definition of the Miwa shift
  is used here for the sake of convenience and is related to the more
  common one by $\t+\mshift{x}=\t-[-x]$.}
$$
\t+\mshift{\lambda}=\left(t_1+\lambda,t_2-\frac{\lambda^2}{2},t_3+\frac{\lambda^3}{3},\ldots,t_i-\frac{(-\lambda)^i}{i},\ldots\right).
$$
Similarly, it is known to satisfy other quadratic difference equations that are more than 3-terms long.  These difference equations are known collectively as the additive formulas \cite{Sato} or the Higher Fay identities \cite{AdlervanMoerbeke}.   
Moreover, any solution to \eqref{HBDE} is necessarily a tau-function of the KP hierarchy \cite{Miwa,Willox}.  Since it is the case that if $\tau(\t)$ satisfies \eqref{HBDE}, it must also satisfy all of the longer difference equations as well\footnote{In keeping with the philosophy of this paper that the HBDE and its fundamental nature can best be understood without reference to more sophisticated results of soliton theory, we wish to point out that the recent paper by Duzhin \cite{Duzhin} can be used to prove that the 3-term relation \eqref{HBDE} implies all of the longer difference equations in an elementary and entirely algebraic way.}, we will focus our attention primarily on this equation.

In the literature, the fact that these equations are satisfied by KP
tau-functions is generally proved as a consequence of higher level
results of soliton theory.  For instance, it can be derived from an application of Wick's theorem to the representation of
tau-functions in terms of the algebra of fermion operators
\cite{Miwa} or through an asymptotic expansion of an integral
equation known to be satisfied by tau-functions
\cite{Willox,Zabrodin}.  

However, a recent trend in the theory of integrable systems is to reconsider difference equations themselves as being fundamental.  In fact, there as been renewed interest in the HBDE \eqref{HBDE} for  its relationship to quantum field theories and in relating quantum to classical integrable systems \cite{KricheverEtAl,Zabrodin}.  
In keeping with this trend, we find it useful to describe the HBDE not as a consequence of the analytic theory of the KP hierarchy, but as a natural consequence of the algebraic geometry of the Grassmannian itself.

\subsection{Grassmann Cone Preserving Maps}\label{sec:gcp}

If $\hat L$ is a linear map from $\bigwedge^{k}\C^n$ to
$\bigwedge^{k'}\C^{n'}$ ($k'\leq k$ and $n'\leq n$)  it is natural to ask whether it preserves the
Grassmann cones.  We will call such a linear map $\hat L$ a \textit{Grassmann
Cone Preserving Map} (or \textit{GCP map}) if it has the property 
$$\hat L(\Gamma^{k,n})\subset
\Gamma^{k',n'}.
$$

As it turns out, it is easy to characterize the
linear maps $\hat L$ which preserve the Grassmann cones in this way.  The
GCP maps are
precisely the ones which have a natural geometric interpretation in
terms of the Pl\"ucker embedding, as we will explain in greater
detail below.  Our description differs from standard treatments of
this question (e.g. \cite{GriffithsHarris}) mainly in that we have
chosen to work with the Grassmann cones rather than projective Grassmannians
to allow us to work with linear rather than rational maps.

First we note that a non-singular linear map $M:\C^n\to\C^{n}$  will
naturally induce a linear GCP map $\hat M:
\bigwedge^k\C^n\to\bigwedge^k\C^n$.
 The map $\hat M$ clearly preserves the Grassmann cone
$\Gamma^{k,n}$ since the image of the decomposable element
$v_1\wedge\cdots\wedge v_k$ is simply $Mv_1\wedge\cdots\wedge Mv_k$.
  In fact, it provides an isomorphism of the
Grassmann cones.  (This equivalently can be interpreted as the
selection of an alternative choice of coordinates for the same
Grassmannian in terms of a different basis of the underlying vector space.)

Another, similar type of linear map on the wedge space that preserves the
Grassmann cones is that induced by a projection map.  Let
$P:\C^n\to \C^{n'}$ be a projection map (i.e. $P^2(v)=P(v)$) and note that the map $\hat
P:\bigwedge^k\C^n\to\bigwedge^k\C^{n'}$ defined by
$$
\hat P(e_{i_1}\wedge e_{i_2}\wedge\cdots\wedge e_{i_k})=Pe_{i_1}\wedge
Pe_{i_2}\wedge\cdots\wedge Pe_{i_k}.
$$
Again, it is obvious that this map is GCP by virtue of its
component-wise action.  Note that
$\hat P$ takes the form of a projection map on $\bigwedge^{k}\C^n$
whose kernel is spanned by all decomposable elements having at least
one component in the kernel of $P$.

A different sort of linear map preserving the Grassmann cones can be
constructed through intersection.  Suppose we have a decomposition of
$\C^n$ as $U\oplus V$ where 
$U$ is a
$p$-dimensional subspace with basis $\{u_1,\ldots,u_p\}$.  We consider
a linear map $\hat U:\bigwedge^k\C^n\to \bigwedge^{k-p}V$ whose
action on decomposable elements of the form $\omega=v_1\wedge v_2\wedge \cdots \wedge v_{k-p} \wedge u_1\wedge
u_2\wedge \cdots \wedge u_p$ is
$$
\hat U(\omega)=\overline{v_1}\wedge \overline{v_2}\wedge \cdots \wedge
\overline{v_{k-p}}
$$
(where the overline indicates projection onto $V$) and 
where $\hat U(\omega)=0$ otherwise.  Geometrically, this corresponds
to intersecting the $k$-dimensional subspace $W$ with $V$ and so it is
clear again that $\hat U$ is a GCP map.
(In the case that the subspace $W$ corresponding to
$\omega\in\Gamma^{k,n}$ is such that $W\cap V$ is not
$(k-p)$-dimensional, $\omega$ is in the kernel of the map $\hat U$.)

Finally, the ``dual isomorphism'' of Grassmannians in which a subspace
$W$ is replaced by its orthogonal complement also takes the form of a
linear map $\bigwedge^k\C^n\to\bigwedge^{n-k}\C^n$ preserving the
Grassmann cones.   One way to explicitly describe the action of this map on the point $\omega=v_1\wedge\cdots \wedge v_k\in\Gamma^{k,n}$ is to construct the $(n+k)\times k$ matrix
$$
\left(I|v_1|v_2|\cdots|v_k\right).
$$
Letting $\pi_{i_1,\ldots,i_{n-k}}$ ($1\leq i_1<i_2<\cdots<i_{n-k}\leq
n$) be the determinant of the sub-matrix of columns
$i_1,i_2,\ldots,i_{n-k},n+1,n+2,\ldots,n+k$ gives the Pl\"ucker
coordinates of the corresponding point in the dual Grassmann cone
$\Gamma^{n-k,n}$.  

The important point is that any
linear map $\hat L$ which preserves the Grassmann cones is necessarily made
up of some combination of the four types of GCP maps described above.
Consequently, if one wishes to show that a certain linear map $\hat L$
is GCP, it makes sense to seek a geometric
interpretation of $\hat L$ as described.  Moreover, if one has a linear map
that is known to be GCP, one could seek a geometric
understanding of its action by finding the projection map, change of
coordinate matrix $M$ and intersecting subspace $V$ such that $L$
takes the form of the composition of the corresponding GCP maps.
It is precisely this philosophy which we apply in attempting to
analyze the geometry of the Hirota Bilinear Difference Equation.

\subsection{Why tau-functions satisfy HBDE}

Suppose that $\tau(\t)$ is a tau-function of the KP hierarchy.  Then there exists some point $\omega\in\Gamma$ such that $\tau(\t)=\tau_{\omega}(\t)$ through Sato's construction.
If we define
\begin{equation}
\pi_{ij}=(\lam{j+3}-\lam{i+3})\tau(\t+\mshift{\lam{i+3}}+\mshift{\lam{j+3}})\hbox{
  for }
-2 \leq i <j\leq 1\label{motivatingmap}
\end{equation}
then the HBDE \eqref{HBDE} becomes the Pl\"ucker relation for $\Gamma^{2,4}$ \eqref{gr24}.
Thus, assuming that $\tau$ satisfies the HBDE, \eqref{motivatingmap} defines a GCP map from $\bigwedge$ to
$\bigwedge^2\C^4$. By the remarks of the Section~\ref{sec:gcp},
the  map \eqref{motivatingmap} ought to have some natural geometric interpretation in terms
of the subspaces corresponding to the points in the Grassmannians.

We present that geometric interpretation here in an explicit form as an alternative way to derive the difference equations satisfied by KP tau-functions and to motivate the more general construction to be presented in the following section.  Note that we present this material without proof, although it can always be reconstructed as a special case of Theorem~\ref{thm:main} which is proved below.  In addition, we note that a similar proof appears in a different context in the paper \cite{MunMart}.

Let $\omega=\sum\pi_I e_I\in \bigwedge$ and $\tau_{\omega}(\t)$ be the tau-function \eqref{eqn:taudef2} associated to it by the usual Sato construction.
 Our method of demonstrating that $\tau_{\omega}(\t)$ satisfies difference equations such as 
 \eqref{HBDE}  when $\omega\in\Gamma$ will depend on interpreting the
 ``Miwa shifts'' $\tau_{\omega}(\t)\mapsto\tau_{\omega}(\t+\mshift{\lambda})$
 as linear maps on $\bigwedge$.  
  Its form is simplified when one recognizes
 the Taylor expansion of a logarithm in the expression so as to write
 $\tau_{\omega}(\t+\mshift{x})$ as the coefficient of $e_{0,1}$ in
 $$
 \tilde\omega(\t+\mshift{x})=\exp(\sum (t_i-\frac{(-x)^i}{i})
 \S^i)\omega=(I+x \S)\tilde\omega(\t).
 $$
 By the same reasoning, $\tau_{\omega}(\mshift{x_1}+\cdots+\mshift{x_k})$ is the
 coefficient of $e_{0,1}$ in
 $$
 \tilde\omega=(I+x_1 \S)\cdots(I+x_k \S)\omega.
 $$
Now we will explicitly determine a formula for this
coefficient as a linear expression in the coordinates $\pi_I$ of $\omega$.

Let 
$$
T(x_1,\ldots,x_k)=\left(I+x_1\S\right)\left(I+x_2\S\right)\cdots
\left(I+x_k\S\right)
$$
be the operator on $H$ depending on the complex parameters 
$x_i$ and consider its extension $\hat T=\hat T(x_1,\ldots,x_k)$ on
$\bigwedge$.

For an arbitrary $\omega=\sum_{I\in\I}\pi_I e_I\in
\bigwedge$, we define the new point $\tilde\omega\in\bigwedge$ and the
new coefficients $\tilde\pi_I$ by the formula
$$
\tilde\omega=\hat T\omega=\sum_{I\in\I}\tilde\pi_I e_I.
$$ 
By virtue of linearity there exists a function $f:\I\to\C$ such that
$$
\tilde\pi_{0,1}=\sum_{I\in \I} \pi_I f(I)
.
$$

As it turns out, it is more natural to describe $f(I)$ in terms of the
numbers which do \textit{not} appear in $I$ rather than those which
do.  We therefore define the notation $J_{j_1,\ldots,j_k}$ to be the
multi-index in $\I$ made up of all integers greater than $-k-1$
\textit{other than} the $k$ specified integers $j_1$ through $j_k$:
$$
I=J_{j_1,\ldots,j_k}=\left\{-k,-k+1,\cdots,0,1,2,\cdots\right\}\backslash
\left\{j_1,\ldots,j_k\right\}.
$$

If $I=J_{j_1,\ldots,j_k}$ (for some integers $j_{\alpha}$ satisfying $-k\leq j_1<j_2<\cdots<j_k$) then $f(I)$ is the Schur function
\begin{equation}
f(I)=\frac{\det(x_{\alpha}^{j_{\beta}+k})_{\alpha,\beta=1}^k}{\det(x_{\alpha}^{\beta-1})_{\alpha,\beta=1}^k}.
\label{eqn:f}
\end{equation}
If $I$ does not take this form, then $f(I)=0$.

We wish now to construct a GCP map
$$
\hat L:\bigwedge\to{\bigwedge}^{k}\C^n
$$
depending on the parameters $\t=(t_1,t_2,\ldots)$ and $(\lambda_1,\ldots,\lambda_n)$ such that the ${n\choose k}$ coefficients of $\hat L(\omega)$ are written
in terms of the functions
$\tau_{\omega}(\t+\mshift{\lambda_{i_1}}+\cdots+\mshift{\lambda_{i_k}})$.  The
Pl\"ucker relations \eqref{plucker} for $\Gamma^{k,n}$ will then take the form of
difference equations for $\tau_{\omega}$ which will be satisfied when
$\omega\in\Gamma$ is an element of the Sato Grassmannian.  

 The ``time variables" enter in the usual manner, by the
exponentiated action of powers of $\S$ (cf. \eqref{eqn:exp}).
Note that $\hat E(\t)$ is already a GCP map from $\bigwedge$ to itself
(for each fixed value of the parameters $\t$, that is).

Similarly, let $P_1:H\to H$ be the projection map defined by 
$$
P_1(e_i)=\left\{\begin{matrix}e_i&\hbox{ if }& i\geq -k\cr
0&\hbox{ if }& i<-k\end{matrix}\right.
$$
that projects onto the  subspace spanned by the elements $e_i$ with
$i\geq -k$.  
The image of $\hat P(\bigwedge)$ is contained in the subspace
$
\bigwedge'
$
spanned by elements of the form $e_{J_{j_1,\ldots,j_{k}}}$.

The dual isomorphism $D$ on $\bigwedge'$ has the effect of replacing the infinite wedge product $e_J$ with the finite wedge product $e_I$ where $J=J_{j_1,\ldots,j_k}$ and $I=(j_1,j_2,\ldots,j_k)$.
We follow this by the extension to the wedge space $\hat M$ 
of the change of basis using an infinite matrix $M$ whose $i^{th}$ row is of the form
$$
\left(\begin{matrix}1&\lambda_i&\lambda_i^2&\lambda_i^3&\cdots\end{matrix}\right)$$
if $i\leq n$ and is equal to the $i^{th}$ row of the identity matrix otherwise\footnote{It is often common to associate a function to an element of $H$ by the rule $e_i=z^i$ (cf. \cite{SegalWilson}).  If one does, then multiplication by the infinite Vandermonde matrix $M$ does nothing other than multiplying the functions by  $z^k$ and evaluating the results at $\lambda_i$ (cf. \cite{MunMart}).  This does simplify the present exposition somewhat, but would not suit the generalization we wish to consider later in which $\S$ is replaced by an arbitrary operator.}
and finally   the extension of the projection map 
$$
P_2(e_i)=\left\{\begin{matrix} e_i&\hbox{if}&i<n+k\cr 0 &\hbox{if}&i\geq n+k\end{matrix}\right.
$$

Now, for each fixed value of the parameters $\lambda_i$ we get a GCP
map $\hat L$ defined as the composition of these GCP maps
$$
\hat L:= \hat P_2\circ \hat M \circ D\circ \hat P_1\circ\hat E:\bigwedge\to{\bigwedge}^{k}\C^n.
$$

The key point is that the map $\hat L$ has been constructed so that
the $n\choose k$ Pl\"ucker coordinates of $\hat L(\omega)$ can be written simply
as Miwa shifts of $\tau_{\omega}$.  Specifically, one can verify by
comparison with \eqref{eqn:f} that its coordinates are precisely
\begin{equation}
\hat \pi_{j_1-k,\ldots,j_k-k}=\Delta(\lambda_{j_1},\lambda_{j_2},\ldots,\lambda_{j_k})\tau_{\omega}(\t+\mshift{\lambda_{j_1}}+\cdots+\mshift{\lambda_{j_k}})
\label{eqn:hatpi1}
\end{equation}
 for $1\leq
j_1<j_2<\ldots<j_k\leq n$ where
$$
\Delta(x_1,\ldots,x_m)=\det(x_i^{j-1})_{i,j=1}^m
$$
denotes the usual Vandermonde determinant.

It is then a consequence of the GCP property of $\hat L$ that
KP tau-functions satisfy difference equations.  In particular, if
$\omega\in\Gamma$, the Pl\"ucker coordinates of $\hat L(\omega)$
satisfy the set of Pl\"ucker relations \eqref{plucker} for
$\Gamma^{k,n}$.  Making the substitution \eqref{eqn:hatpi1}, these
algebraic equations in the parameters $\pi_I$ take the form of
difference equations for $\tau_{\omega}$.  For instance, in the case
$k=2$, $n=4$, the Pl\"ucker coordinates \eqref{eqn:hatpi1} satisfy
\eqref{gr24}, which is nothing other than the HBDE \eqref{HBDE}.

\section{KP Generators}\label{sec:KPgen}
\subsection{Preliminaries}
Let $\omega\in\bigwedge$ and let $S:H\to H$ be an unspecified linear operator\footnote{In order to ensure that the operations we utilize will be well defined, we assume that the operator $S$ is bounded and is ``almost lower triangular'', i.e. that it can be written in the block form
$$
S=\left(\begin{matrix}A&0\cr C&D\end{matrix}\right)
$$
with $D$ strictly lower triangular with respect to \textit{some} splitting of the underlying space $H$}.  Define $\tau_{\omega}^{S}(\t)$ again by
\begin{equation}
\tau_{\omega}^{S}(\t)=\tilde\pi_{0,1,2,\ldots}(\t)
\qquad
\tilde\omega(t)=\hat E(\t) \omega=\sum_{I\in\I}\tilde\pi_I(\t) e_I.
\qquad
E(\t)=\textup{exp}\sum_{i=1}^{\infty} t_i S^i,\label{eqn:taudef}
\end{equation}
and call $S$ a \textit{KP Generator} if it has the property that $\tau_{\omega}^{S}(\t)$ is a tau-function whenever $\omega\in\Gamma$.  Our goal is to determine what operators $S:H\to H$ are KP generators.
Of course, we know that $S=\S$ is one such generator.  In addition, $S=0$ provides a trivial example for which $\tau_{\omega}^{S}(\t)$ is constant.  However, as we will see, there is a larger class of generators which produce non-trivial KP tau-functions than just $S=\S$.

We will proceed by attempting to construct a linear GCP map such that the Pl\"ucker coordinates of the image are appropriate Miwa-shifted copies of $\tau_{\omega}^{S}$.  Whether such a map exists depends upon the block decomposition 
\begin{equation}
S=\left(\begin{matrix}S_-\cr S_+\end{matrix}\right)=
\left(\begin{matrix}S_{--}&S_{-+}\cr S_{+-}&S_{++}\end{matrix}\right)
\end{equation}
with respect to the splitting \eqref{H-+H+}.  

Note that it follows from an elementary calculation that there is no harm in conjugating $S$ by a block upper triangular matrix.
\begin{lemma}\label{lem:Gact}
Let $G:H\to H$ be an invertible operator with block decomposition
$$
G=\left(\begin{matrix}A&B\cr0&C\end{matrix}\right).
$$
with respect to the splitting \eqref{H-+H+}, and where $C$ is almost lower unipotent.  Then for any $\omega\in\bigwedge$, the functions $\tau_{\omega}^S(\t)$ and $\tau_{\omega'}^{S'}(\t)$ differ by the constant multiple $\det C$ where $S'=GSG^{-1}$ and $\omega'=\hat G\omega$.  Consequently, $S$ is a KP generator if and only if $S'$ is a KP generator.
\end{lemma}
We make use of this lemma to assume, without loss of generality, that the matrix $S_{++}$ is lower triangular in the remainder of the paper.

\subsection{A Rank One Condition}

We will show that $S$ being a KP generator is equivalent to the following restriction on the rank of the block $S_{+-}:H_-\to H_+$:
\begin{equation}
\textup{rank}(S_{+-})\leq 1.\label{rk1}
\end{equation}
It is notable that there is a long precedent of such ``rank one conditions'' in the literature of integrable systems
(cf. \cite{AdenCarl,BS,CS,TMP,HaineIliev,myBetheAnsatz,KG,Rothstein,RS,sakh,WentingYishen,WilsonCM}.)

\begin{lemma}\label{lem:Lexists}
If $S$ satisfies the rank one condition \eqref{rk1}, then the linear map
$\hat L_{k,n}:\bigwedge \to \bigwedge^{k,n}$ defined by giving the coordinates of $\hat L_{k,n}(\omega)$ the values
\begin{equation}
\hat\pi_{i_1-k,\ldots,i_k-k}=\Delta(\lambda_{i_1},\ldots,\lambda_{i_k})\tau_{\omega}^{S}(\t+\mshift{\lambda_{i_1}}+\cdots+\mshift{\lambda_{i_k}}),\label{eqn:hatpi2}
\end{equation}
 (for $1\leq i_1<i_2<\cdots<i_k\leq n$) is a Grassmann Cone Preserving (GCP) map.
\end{lemma}
\begin{proof}
Due to linearity it is sufficient to assume that $\omega$ is an elementary wedge product.  So, we suppose  $\omega=v_1\wedge v_2\wedge\cdots$ and throughout the remainder of the proof we will consider $\omega$ to be an infinite matrix whose $i^{th}$ column is the representation of $v_i$ in the basis $\{e_i\}$.
In addition, we note at this point that it is sufficient to prove the claim for $\t=(0,0,0,\ldots)$ but arbitrary $\omega$  since $\tau_{\omega}^{S}(\t)= \tau_{\omega'}^S(0,0,\ldots)$ if $\omega'=\tilde\omega(\t)$.

We will show that there exist operators $A$ and $M$ on the underlying vector space such that the map $\hat L_{k,n}$ can be decomposed into 
the composition of the change of basis and projection
$\hat A$ followed successively by the dual isomorphism, the change of
basis $\hat M$, and the map induced by the orthogonal
projection onto the subspace spanned by $\{e_{-k},\ldots,e_{n-k-1}\}$.  The GCP nature of $\hat L_{k,n}$ is then clear by virtue of the fact that each of these component maps is GCP.

Denote
$$ P(x) = (1+ x_1 x)\cdots (1 + x_k x)= \sum_{i=0}^k \sigma_i x^i\ ,$$
where 
$$
\sigma_i(x_1,...,x_k) = \sum_{1\le \alpha_1 < \ldots < \alpha_i \le k} 
x_{\alpha_1} \cdots x_{\alpha_i}
$$
is the $i$th elementary symmetric function of $x_1,...,x_k$.

Then, $\tau_{\omega}^{S}(\mshift{x_1}+\cdots+\mshift{x_k})$ is the ``first" Pl\"ucker coordinate of $P(S)\omega$ and hence
$$
\tau_{\omega}^{S} (\mshift{x_1} + \ldots \mshift{x_k})= \det ( P(S)_+ \ \omega). 
$$
But, since $S_{+-}$ has rank one, there exist vectors $u$ and $v$ such that
$S_{+-}=uv^{T}$.  Then
$$
{\displaystyle
\begin{array}{rl} P(S)_+ =&  \sum_{i=0}^k \sigma_i (S^i)_+ \\ & \\
= & \left ( \begin{array} {cc} 0 & P(S_{++}) \end{array} \right )
+ \sum_{i=0}^k \sigma_i \sum_{j=0}^{i-1} (S_{++})^{i-j-1} u v^T (S^j)_-\\ &\\
= & \left ( \begin{array} {cc} 0 & P(S_{++}) \end{array} \right )
+ \sum_{j=0}^{k-1}  \sum_{i=j+1}^{k } \left ( \sigma_i\, (S_{++})^{i-j-1} u\right )
\left ( v^T (S^j)_- \right )\\ &\\
=& P(S_{++}) \left (\left [ \begin{array} {cc} 0 & I \end{array} \right ] 
+  \sum_{j=0}^{k-1} \sum_{r=1}^{k} c_{r j} ( I + x_r S_{++})^{-1} u  \left ( v^T (S^j)_- \right )  \right )\ ,
\end{array} 
}
$$
where $c_{rj}$ are the coefficients in a partial fractions decomposition
$${\displaystyle
\frac{\sum_{i=j+1}^{k } \sigma_i x^{i-j-1}} {P(x)} = 
\sum_{r=1}^{k} \frac{c_{rj}}{1 + x_r x}\ .
}
$$
More explicitly,
$${\displaystyle
c_{rj} = x_r^k \frac{ \sum_{\alpha=0}^{j} \sigma_\alpha (- x_r)^{j-\alpha}}
{\prod_{s\ne r} (x_r -x_s)}\ .
}
$$
Note also, that if we denote by $\sigma_\alpha^r$ the $\alpha$th elementary
symmetric function in $x_1,\ldots, x_{r-1}, x_{r+1}, \ldots, x_k$, then
$${\displaystyle
\begin{array}{c}
\sum_{\alpha=0}^{j} \sigma_\alpha (- x_r)^{j-\alpha}= \sum_{\alpha=0}^{j} (\sigma^r_\alpha + x_r \sigma^r_{(\alpha-1)}  )(- x_r)^{j-\alpha}\\
= \sigma_j^r\ .
\end{array}
}
$$
Thus,
$$
c_{rj} = x_r^k \frac{ \sigma_j^r}
{\prod_{s\ne r} (x_r -x_s)}\ .
$$

Next, denote by $U$ a matrix with columns
$( I + x_r S_{++})^{-1} u, \ r=1,\ldots, k$, by $V$ a matrix
with rows $v^T (S^{k-j})_-, \ j=1,\ldots k$, and by
$C$ the matrix $ (c_{r,k-j})_{r,j=1}^k$.
Observe that
$$
C = \mbox{diag} (x_1^k,\ldots, x_k^k) \mbox{Van} (x_1,\ldots, x_k)^{-1}\ , 
$$ 
where $\mbox{Van} (x_1,\ldots, x_k)=(x_j^{r-1})_{r,j=1}^k $ is the Vandermonde matrix of 
$(x_1,\ldots, x_k)$. In particular
$$
\det C = \frac{ x_1^k \cdots x_k^k} {\Delta (x_1,\ldots, x_k)}, 
$$
where, again, $\Delta (x_1,\ldots, x_k)=\det \mbox{Van} (x_1,\ldots, x_k)$.

We see now that
$$
{\displaystyle
P(S)_+ = P(S_{++}) \left (\left [ \begin{array} {cc} 0 & I \end{array} \right ] 
+  U C V \right )
}
$$
and 
$$
\tau_{\omega}^{S} (\mshift{x_1} + \ldots \mshift{x_k})= \det ( P(S)_+ \ \omega) = \det P(S_{++})
\det ( \omega_+ + U C V\omega)\ ,
$$
where we used a natural decomposition 
$\omega=\left (\begin{array}{c} \omega_-\\ \omega_+\end{array} \right )$.
Using the Schur complement formula for determinants of $2\times2$ block
matrices with square diagonal blocks:
$$
\det \left | \begin{array}{cc} Z_{11} & Z_{12}\\  Z_{21} & Z_{22}\end{array} \right |
=\det Z_{11} \det ( Z_{22} - Z_{21}Z_{11}^{-1} Z_{12})\ ,
$$
we obtain
$$
\det ( \omega_+ + U C V\omega) = (-1)^k \det C 
\det \left | \begin{array}{cc} - C^{-1} & V\omega\\  U  & \omega_+\end{array} \right |\ .
$$
Noting that $\det P(S_{++}) = \det ( I + x_1 S_{++}) \cdots \det ( I + x_k S_{++})$,
we finally conclude that
\begin{equation}
\tau_{\omega}^{S} (\mshift{x_1} + \ldots \mshift{x_k})= \frac{1}{\Delta (x_1,\ldots, x_k)}
\det \left ( f(x_1)| ... | f(x_k)| (A\omega)  \right ) \ ,
\label{bleh}
\end{equation}
where $f(x)$ is a column vector of the form
$$
f(x)= \mbox{col} (p_0(x), -x p_0(x), \ldots, (-x)^{k-1} p_0(x), (-x)^k p_1(x), (-x)^k p_2(x), ...)
$$
with
$$ p_0(x)= \det ( I + x S_{++}), \ (p_i(x))_{i\ge 1} = p_0 (x) ( I + x S_{++})^{-1} u $$
and
$$
A=\left ( \begin{array}{cc} V_- & V_+ \\  0 & I\end{array}\right ) ,
$$
where we re-wrote  $ V = \left ( \begin{array}{cc} V_- & V_+ \end{array}\right )$.

We can construct a corresponding GCP map $\hat L_{k,n}$ as follows.  The matrix $A$ above can be viewed as a combination of a projection and a change of coordinates, and so its extension $\hat A$ to the wedge space takes the form of a GCP map.  Moreover, if we define the infinite matrix $M$ to be the matrix whose $i^{th}$ row is $f(\lambda_i)^T$  for $i\leq n$ and is the $i^{th}$ row of the identity matrix otherwise, then $\hat M$ is a GCP map which represents the change in coordinates corresponding to an alternative choice of underlying basis.

The claim then follows from recognizing \eqref{bleh} as the statement that
$$
\Delta(\lambda_{i_1},\ldots,\lambda_{i_j})\tau(\mshift{\lambda_{i_1}}+\cdots+\mshift{\lambda_{i_k}})
$$ is the minor of the matrix $[f(\lambda_1)\ \cdots f(\lambda_n)\ \ A\omega]$ in which all rows and all
columns from column $n+1$ onwards are chosen, but only columns
$i_1,\ldots,i_k$ from the first $n$ are selected and noting that this and consequently can be interpreted as the composition of the dual isomorphism with the map $\hat M$. 
\end{proof}

By the GCP nature of the map, we can use the Pl\"ucker relations to determine equations satisfied by $\tau_{\omega}^{S}$ when $\omega\in\Gamma$.
\begin{corollary}\label{cor1}
Define $\tau_{\omega}^{S}$ by \eqref{eqn:taudef}.  Then if $\omega\in\Gamma$ and $S_{+-}$ is an operator of rank one, $\tau_{\omega}^{S}$ satisfies a collection of difference equations obtained by substituting \eqref{eqn:hatpi2} into \eqref{plucker}.
\end{corollary}

Conversely, we conclude that no such map exists in the case $k=2$, $n=4$ if the operator $S$ does not have the rank one property.
\begin{lemma}
It the operator $S$ does not satisfy the rank one condition \eqref{rk1}, then
the linear map  $\hat L_{2,4}:\bigwedge \to \bigwedge^{2,4}$ defined by the property that $\hat L(\omega)$ has coordinates
$$
\hat\pi_{i_1-3,i_2-3}=(\lambda_{i_2}-\lambda_{i_1})\tau_{\omega}^{S}(\t+\mshift{\lambda_{i_1}}+\mshift{\lambda_{i_2}})
$$
(for $1\leq i_1<i_2\leq 4$) is \textit{not} a GCP map.
\end{lemma}
\begin{proof}
It suffices to show that there exists a point $\omega\in\Gamma$ and values for the parameters $\lambda_1,\ldots,\lambda_4$ such that $\hat L_{2,4}(\omega)\not\in \Gamma^{2,4}$ at $\t=0$.  We will show, in particular, that if $S$ does not satisfy the rank one condition then it is possible to find an $\omega\in\Gamma$ such that$$
\hat L_{2,4}(\omega)=\pi_{-2,-1}e_{-2,-1}+\pi_{0,1}e_{0,1}\qquad \pi_{-2,-1}\not=0,\pi_{0,1}\not=0.
$$

For notational convenience, 
we will denote by $LT$  the unspecified lower triangular entries of various matrices.
Thus, since $S_{++}$ is lower triangular we can state that
$$ S_{++} = \sum_{1\le i \ll \infty} s_i E_{ii} + LT\ $$
and since $S_{+-}$ is not of rank one, it has to have the form
$$ S_{+-} = (e_{i_1} + \sum_{i=i_1+1}^{i_2-1} c_i e_i) v_1^T +
e_{i_2}  v_2^T  +  \sum_{i>i_2} e_i v_i^T\ ,$$
where $1\le i_1 < i_2$ and vectors  $v_1, v_2$ are linearly independent.

Choose vectors $w_1, w_2$ such that $v_{j}^T w_j = d_j \ne 0\ (j=1,2) $ and
$v_{1}^T w_2 = 0$. Then
$$ S_{+-} ( w_1 e_{i_1}^T +  w_2 e_{i_2}^T) = d_1 E_{i_1 i_1} + d_2 E_{i_2 i_2} + LT\ .$$
For $\mu \in\mathbb{C}$, define $\omega=\omega(\mu)=\left (\begin{array} {c} \omega_-\\ I\end{array}\right )$
with
$$ \omega_- = \mu ( w_1 e_{i_1}^T +  w_2 e_{i_2}^T)\ .$$
Then
$$ \left ( ( \lambda S + I)\omega\right )_+ = \sum_{ i \ne i_1, i_2} ( \lambda s_i+1)  E_{ii} + \sum_{ j=1}^2 ( \lambda (s_{i_j}+ \mu d_j)+1)  E_{i_j i_j} + LT\ .$$
Clearly,
$\tau_\omega^S (\mshift{\lambda}) = \det \left ( ( \lambda S + I)\omega\right )_+$ is not identically zero, but
$\tau_\omega^S (\mshift{\lambda_j}) = 0$ for $ \lambda_j =\lambda_j(\mu)= \frac{-1}{s_{i_j}+ \mu d_j} $ ($j=1,2$)
 where the constants $d_j$ should be selected in such a way that two linear functions of $\mu$, $s_{i_j}+ \mu d_j$ are not identically equal.

Observing that
$$
\lim_{\mu \to \infty} \left ( ( \lambda_1 S + I) ( \lambda_2 S + I)\omega\right )_+ = \sum_{ i \ne i_1, i_2}   E_{ii} - \frac{d_1}{d_2}  E_{i_1 i_1} -  \frac{d_2}{d_1}  E_{i_2 i_2} + LT\ ,
$$
we conclude that
$$
\lim_{\mu \to \infty} \tau_{\omega(\mu)}^S (\mshift{\lambda_1(\mu)}+ \mshift{\lambda_2(\mu)})=1.
$$

Then, 
 there exists $\mu$ such that for $\omega=\omega(\mu)$, $\lambda_1=\lambda_1(\mu)$, $\lambda_2=\lambda_2(\mu)$, $\lambda_4=0$
and almost every $\lambda_3$
$$
\tau_\omega^S (\mshift{\lambda_j})=0\ (j=1,2)\ , \ \tau_\omega^S (\mshift{\lambda_3})\ne 0\ \mbox{and}\
\tau_\omega^S (\mshift{\lambda_1}+ \mshift{\lambda_2})\ne 0\ .
$$

\end{proof}

Combining the two lemmas above, and using the equivalence of the difference equations of Corollary~\ref{cor1} to the KP hierarchy \cite{Miwa,Sato}, we conclude that
\begin{theorem}\label{thm:main}
The function $\tau_{\omega}^{S}(\t)$  \eqref{eqn:taudef}  is a tau-function of the KP hierarchy for all $\omega\in\Gamma$ if and only if $S$ has a decomposition \eqref{H-+H+} such that $S_{+-}$ satisfies the rank one condition \eqref{rk1}.
\end{theorem}

\subsection{Characterizing Grassmannians using KP}

The functions $\tau_{e_I}^S(\t)$ for $I\in\I$ play an important role.
By linearity, we see that for arbitrary $\omega\in\bigwedge$ the
function $\tau_{\omega}^S(\t)$ can be expanded as a sum
\begin{equation}
\tau_{\omega}^S(\t)=\sum_{I\in\I}\pi_I \tau_{e_I}^S(\t)\label{eqn:sum}
\end{equation}
 where $\pi_I$ are the Pl\"ucker coordinates of $\omega$
($\omega=\sum \pi_I e_I$).
Then, by virtue of the main result of the previous section, we can say that if $S$ satisfies \eqref{rk1} then the linear combination 
\eqref{eqn:sum}
is a tau-function if the coefficients $\pi_I$ are the Pl\"ucker coordinates of a point $\Gamma$.

In the case $S=\S$, this is the well-known decomposition of the
tau-function into a sum of Schur polynomials
\cite{Sato,SegalWilson}.  However, in that case there is
something stronger one can say.  In the standard construction one also
has that the linear combination \eqref{eqn:sum} is a tau-function
\textit{only} if the coefficients are chosen to be the coordinates of
a point in $\Gamma$.  In this way, the standard Sato construction
provides a way to determine whether a given $\omega=\sum \pi_I e_I$
lies in the Grassmann cone via the KP hierarchy.
This is not the case for every $S$ selected to satisfy \eqref{rk1}.
In order to be able to say that $\tau_{\omega}^S(\t)$ is a
tau-function \textit{only} if $\omega$ lies in a (finite) Grassmann
cone additional restrictions will have to be placed on the selection of $S$.

We say that the the KP generator $S:H\to H$ satisfying \eqref{rk1}  is  \textit{$(k,n)$-faithful} if the function
$$
\sum_{I\in\I_{k,n}} \pi_I\tau_{e_I}^S(\t)
$$
is \textit{not} a tau-function of the KP hierarchy 
when 
$$
\omega=\sum_{I\in\I_{k,n}}\pi_I e_I
$$
lies outside of the Grassmann cone $\Gamma^{k,n}\subset\bigwedge$.
Similarly, we will say that $S$ is \textit{faithful} if the function \eqref{eqn:sum}
is a tau-function of the KP hierarchy only for $\pi_I$ that are coordinates of a point in $\Gamma$.
Note that if $S$ is $(k,n)$-faithful then it is necessarily $(k',n')$-faithful for $k'\leq k$ and $n'\leq n$ and that it is $(k,n)$-faithful for any choice of $k<n$ if it is faithful.

\begin{lemma}
Let $S:H\to H$ satisfying \eqref{rk1} and let $K\subset \bigwedge$ be the subspace
$$
K=\left\{\omega\in\bigwedge:\omega=\sum_{I\in\I_{k,n}}\pi_I e_I,\ \hat L_{k,n}(\omega)\equiv0\right\},
$$
where $\hat L_{k,n}$ is the linear map defined in Lemma~\ref{lem:Lexists}.
Then $S$ is $(k,n)$-faithful if and only if $K=\{0\}$.  Consequently, $S$ is faithful if $\tau_{\omega}^S(\t)\equiv0$ only for $\omega=0\in\bigwedge$.
\end{lemma}
\begin{proof}
Suppose $\omega'\in K$ has the property that $\hat L_{k,n}(\omega')=0$ for all values of the parameters.  This means that $\tau_{\omega+\omega'}^S(\t)$ is a tau-function whenever $\tau_{\omega}^S(\t)$ is a tau-function.  The only point in $\omega'=\Gamma^{k,n}$ which has the property that $\omega'+\Gamma^{k,n}=\Gamma^{k,n}$ is $\omega'=0$, and so if $S$ is $(k,n)$-faithful then $K=\{0\}$.  
On the other hand, if $K=\{0\}$ then $\hat L_{k,n}$ gives an isomorphism of $\Gamma^{k,n}\subset\Gamma$ with $\Gamma^{n-k,n}$ such that the difference equations satisfied by $\tau_{\omega}^S(\t)$ are precisely the Pl\"ucker relations.    That these conditions are satisfied for all  $k<n$ is equivalent to confirming that $\tau_{\omega}^S(\t)$ is never the zero function if $\omega\not=0$.
\end{proof}

Clearly, one requirement for faithfulness which is not imposed by \eqref{rk1} is that the powers of $S$ followed by projection onto $H_+$ cannot all be trivial for any element of $H_-$; otherwise that element would be ``invisible'' to the procedure for producing tau-functions.

\begin{theorem}
If $S$ is $(k,n)$-faithful then for $v\in\langle e_{k-n},\ldots,e_-1\rangle$ there is some $m$ ($1\leq m\leq k$) such that $S^m v\not\in H_-$.  If $S$ is faithful then for $v\in H_-$ there is some $m$ such that $S^m v\not\in H_-$.
\end{theorem}
\begin{proof}
If no power of $S$ applied to $v\in H_-$ results in a positive projection onto $H_+$ then for $\omega=v\wedge e_1\wedge e_2\wedge\cdots$ the coefficient of $e_{0,1}$ in $\hat S^j \omega$ will always be zero and then $\tau_{\omega}^S(\t)\equiv 0$, which implies that $S$ is not faithful.  

More specifically, the conditions given for $(k,n)$-faithfulness correspond to the non-singularity of the matrix $A$ which appears in the proof of Lemma~\ref{lem:Lexists}.  If these conditions are not met then the GCP induced by $A$ will have a non-trivial kernel, preventing $S$ from being faithful according to the previous lemma.\end{proof}

\section{Applications}\label{sec:Appl}
\subsection{Symmetries}

There are several obvious group actions on the set of operators $S$ satisfying the rank one condition \eqref{rk1}.  These translate into
symmetries of the KP hierarchy through the function
$\tau_{\omega}^S(\t)$.  

For instance, consider the fact that the set of solutions to \eqref{rk1} is closed under scalar multiplication.   If we define the scalar multiple of $\t=(t_1,t_2,\ldots)$ by
$$
\lambda\t=(\lambda t_1,\lambda^2 t_2,\lambda^3 t_3,\ldots)
$$ then the ``scale invariance" of the KP hierarchy is represented by
the fact that $\tau(\lambda\t)$ is a KP tau-function whenever
$\tau(\t)$ is one (for $0\not=\lambda\in\C$).  This can be easily
verified by noting that
$$
\tau_{\omega}^{\lambda S}(\t)=\tau_{\omega}^S(\lambda\t).
$$

Also for $\lambda\in\C$, we see that $S+\lambda I$ satisfies the rank
one condition whenever $S$ does.  The result is a translation of the
time parameters similar to the ``Miwa shift'' described earlier:
$$
\tau_{\omega}^{S+\lambda I}(\t)=\tau_{\omega}^S(t'_1,t'_2,t'_3,\ldots)
$$
with
$$
t'_j=\sum_{i=0}^{\infty} {i+j\choose i}\lambda^i t_{i+j}.
$$

Other symmetries are manifested as a change in the choice of $\omega$
rather than as a function of the parameters $\t$.
If $G:H\to H$ is an operator satisfying the conditions of Lemma~\ref{lem:Gact} then $S'=GSG^{-1}$ satisfies the rank one condition whenever $S$ does and
$$\tau_{\omega}^{S'}(\t)=\det(C)\tau_{\hat G\omega}^S(\t).$$

\subsection{Finite Grassmannians and Rank One Conditions}

Let $\omega\in\Gamma$ be chosen so that the only non-zero Pl\"ucker
coordinates are those with multi-indices in $\I_{k,n}$ (so we can
consider $\omega$ as being an element of $\Gamma^{k,n}$).  If $S$ is
chosen to have the an appropriate block lower triangular
structure\footnote{ Consider the decomposition of $H$ into $$
H=H_{<}\oplus H_{0}\oplus H_{>}
$$ where $H_{<}$ is spanned by the basis elements $e_i$ for $i<k-n$,
$H_{0}$ is spanned by $e_i$ with $k-n\leq i\leq k-1$ and $H_{>}$ is
spanned by $e_i$ with $i\geq k$ and the corresponding decomposition of
$S$ into
\begin{equation}
S=\left(\begin{matrix}S_{<<}&S_{<0}&S_{<>}\cr S_{0<}&S_{00}&S_{0>} \cr  S_{><}&S_{>0}& S_{>>}\end{matrix}\right).\label{otherblock}
\end{equation}
Then the property of having only zero Pl\"ucker coordinates for
 $I\not\in\I_{k,n}$ is preserved by the flows generated by $S$ as long
 as $S_{<0}=S_{<>}=S_{0>}=0$.  If only $S_{>>}$ is also equal to zero,
 then it is sufficient to consider simply an $n\times n$ matrix $S$
 generating flows on the finite Grassmannian $\Gamma^{k,n}$.}  then this property is conserved and
 the flows generated by powers of $S$ are all contained in the finite
 dimensional Grassmannian $\Gamma^{k,n}$.  In the standard
 construction with $S=\S$, this necessarily produces tau-functions
 which are polynomials in the variables $t_1,\ldots,t_n$ since $\S$ is
 nilpotent on $\C^n=\langle e_{k-n},\ldots,e_k\rangle$.  However, if
 we are willing to consider more general $S$, then other solutions can
 be constructed from flows on finite dimensional Grassmannians as
 well.

One special class of solutions of the KP hierarchy are those coming
from the Grassmannian $Gr^{rat}$ \cite{WilsonCM}, i.e. those whose
algebro-geometric spectral data are a line bundle over a (singular)
\textit{rational} spectral curve.  This class of solution includes the
rational solutions and the soliton solutions as well as other
solutions which can be written using exponential and rational
functions.   As we will see, these are the only solutions which can be
obtained in the case of a finite dimensional Grassmannian regardless
of the choice of KP generator $S$. 

Let $\tau(\t)$ be such a tau-function and associate to it the
``stationary wave function'' $\psi(x,z)=\tau(\vec
x-[z^{-1}])e^{xz}$ ($\vec x=(x,0,0,\ldots)$). The solutions in
$Gr^{rat}$ can be identified by two pieces of data: a polynomial
$p(z)$ of degree $n$ such that $p(z)\psi(x,z)$ is non-singular in $z$
and an $n$-dimensional space of finitely supported distributions in $z$ that annihilate
$p(z)\psi(x,z)$.  The tau-function can then be written conveniently in
a Wronskian form utilizing these distributions \cite{SegalWilson} and
viewed as coming from a flow on a finite dimensional ``dual''
Grassmannian \cite{CMBis}.

It is not difficult to see (cf. Theorem 2 in \cite{TMP}) that in the case of a KP generator having
the block decomposition specified in the footnote and with a $k\times
n$ matrix $C$ representing $\omega\in \Gamma^{k,n}$, the stationary
wave function $\psi(x,z)$ takes the form
$$
\psi(x,z)=\frac{\det\left([0\ I] e^{x
    S_{00}}\left(zI-S_{00}\right)C\right)}{p(z)\det([0\ I]e^{x
    S_{00}}C)}e^{xz}.
$$
We then see that this is a solution in $Gr^{rat}$ for which 
$p(z)=z^n+O(z^{n-1})$ is a polynomial depending on the block $S_{>>}$
(it is just $z^n$ in the case $S_{>>}=0$)
and the distributions have support at the eigenvalues of the finite
block $S_{00}$ (with degrees bounded by the multiplicities).

It is not a coincidence that both rational solutions and soliton
solutions have been frequently described in terms of ``rank one
conditions'' on finite matrices in the literature of integrable
systems \cite{BS,TMP,myBetheAnsatz,KG,Rothstein,RS,WilsonCM}.  These
rank one conditions are merely special cases of the more general
constraint \eqref{rk1} as seen in the following examples.

Let $X$, $Y$ and $Z$ be $n\times n$ matrices 
and 
consider the case in which $S$
has the block form
$$ S=\left(\begin{matrix} Z&0\cr XZ-YX& Y\end{matrix}\right).
$$
Then if  $\omega=v_1\wedge\cdots\wedge v_n\in \Gamma^{n,2n}$ is chosen so that $v_i^T$ is the $i^{th}$ row of the matrix $(I\ I+X)$
one finds that
$$
\tau_{\omega}^S(\t)=\det\left(\exp\left(\sum_{i=1}^{\infty} t_i Z^i\right) X + \exp\left(\sum_{i=1}^{\infty} t_i Y^i\right)\right).
$$ 
It is known that this formula gives a tau-function of
the KP hierarchy precisely when the matrix $XZ-YX$ has rank one \cite{KG},
but as this happens to be the lower-left block of the matrix $S$ we
can now also see this as a consequence of Theorem~\ref{thm:main}.

The matrices $X$, $Y$ and $Z$ can be selected so as to make $\tau_{\omega}^S(\t)$ the tau-function of an $n$-soliton solution to the KP hierarchy\footnote{It has already been noted in other contexts that
$n$-soliton solutions ``live" in finite dimensional Grassmannians
\cite{CMBis,Kodama}.}
by choosing $4n$ complex parameters $\mu_i$, $\lambda_i$, $\alpha_i$
and $\gamma_i$  ($1\leq i\leq n$, such that
$\mu_i\not=\lambda_j$) 
and letting
$$
X_{i,j}=\frac{\alpha_i}{\beta_j(\lambda_j-\mu_i)}, \ Y_{i,j}=\mu_i\delta_{ij},\ 
Z_{i,j}=\lambda_i\delta_{ij}.
$$

Similarly, if $X$ and $Z$ are $n\times n$ matrices which satisfy the
``almost-canonically conjugate" equation $$\textup{rank}(XZ-XZ+I)=1$$
then it is known that $$\tau(\t)=\det\left(X+\sum_{i=1}^{\infty} i t_i Z^{i-1}\right)$$ is a
tau-function whose roots obey the dynamics of the Calogero-Moser
Hamiltonian \cite{WilsonCM}.  This too can be seen as a special case of the selection of an appropriate KP generator satisfying the rank one condition \eqref{rk1} where
$$
S=\left(\begin{matrix}
Z&0\cr
XZ-ZX+I&Z
\end{matrix}\right)
$$ 
and $\omega\in\Gamma^{n,2n}$ is chosen as in the last example.

There is interest in other special subclasses of solutions from
$Gr^{rat}$,
such as positon, negaton and complexiton solutions.  Without
going into details, we note that these kinds of solutions can be
obtained by selecting a finite-dimensional KP generator $S$ with an
appropriate spectral structure. In particular, $S$ can be a real
$N\times N$ upper Hessenberg (upper triangular plus lower shift)
matrix with a prescribed characteristic polynomial.  Then, for any
$k$, $S$ satisfies then rank one condition with respect to the
splitting $\mathbb{C}^N= \langle e_1, \ldots, e_k\rangle \oplus
\langle e_{k+1}, \ldots, e_N\rangle$.  For example, if $S$ is chosen
to have complex eigenvalues, then, for any real $\omega$ in
$\Gamma^{k,N}$, $\tau_\omega^S$ is a real KP tau-function of a
complexiton type.


\subsection{Discrete KP (dKP)  hierarchy}

This hierarchy of differential-difference equations is described by equations \eqref{eqn:Laxopr}, \eqref{eqn:KPhier}
with $\partial$ replaced with the difference operator $D$ ( $(D f)(k)= f(k+1) - f(k)$ )
and $w_i(\t)$ replaced with multiplication operators ( $(w_i(\t) f)(k)= w_i(k;\t) f(k)$ )
acting on functions of a discrete variable $k\in \mathbb{Z}$ 
(see, e.g., \cite{HaineIliev}). Similarly to the continuous case, the solution
has a form $\calL:=W\circ\partial\circ W^{-1}$ with $W$ constructed from a dKP
tau-function $\tau(k;\t)$:
$$
W=\frac{1}{\tau}
\tau(t_1-D^{-1},t_2-\frac12D^{-2},\ldots)\ .
$$
It was shown in \cite{HaineIliev}, that if $\tau(\t)$ is a tau-function
for the continuous KP hierarchy, then
$$
\tau(k;\t)=\tau(t_1+k, t_2 -\frac{k}{2}, t_3 + \frac{k}{3},\ldots) = \tau(\t +  k \mshift{1})
$$
is a dKP tau-function. Then Theorem~\ref{thm:main} implies

\begin{corollary}
If $S$ has a decomposition \eqref{H-+H+} such that $S_{+-}$ satisfies the rank one condition \eqref{rk1}, then, for any $\omega\in\Gamma$,
$$ \tau_{\omega}^{S}(k;\t) = \det \left( ( (I + S) ^k \tilde\omega(\t))_+\right )
$$
is a tau-function of the dKP hierarchy.
\end{corollary}

Taking a limit as  $x_1,\ldots, x_k \to 1$ in \eqref{bleh}, one obtains
a Wronskian representation for $\tau_{\omega}^{S}(k;\t)$:
$$
\tau_{\omega}^{S}(k;\t) = \det \left ( f(1)| f'(1)|... | f^{(k)}(1 )|
(A \hat E(\t)\omega)  \right ) \ .
$$

\subsection{Singularities}

The Lax operator $\calL$ has a singularity wherever the corresponding tau-function has a zero.  This clearly happens at $\t=0$ if and only if the corresponding point in the Grassmannian is outside of the ``big cell'' \cite{SegalWilson}.  Moreover, the degree of this singularity has been related to more specific information about the location of the corresponding point in the Grassmannian for the standard construction with $S=\S$ \cite{AdlervanMoerbeke}.  A similar result is a necessary consequence of the rank one condition \eqref{rk1} for more general choices of $S$ as well.

Consider the subset of $\Gamma$ of elements that can be written as a wedge product with sufficiently many components in $H_-$
$$
\Gamma_{k}=\left\{\omega\in\Gamma| \omega=v_1\wedge v_2 \wedge \cdots,\ v_i\in H_-\ \hbox{for }1\leq i\leq k\right\}.
$$
If $\omega\in\Gamma_{k}$ for $k>0$ then $\tau_{\omega}^S(0)=0$ regardless of whether $S$ satisfies the rank one condition \eqref{rk1}.    In general, whether the result of a single Miwa shift,  $\tau_{\omega}^S(0+\mshift{\lambda})$, is non-zero depends on the choice of $S$ regardless of $k$.  However, as the following result shows, if $S$ is selected to satisfy the rank one condition \eqref{rk1} then at least $k$ Miwa shifts are required to get a non-zero value for the tau-function if $\omega$ is in $\Gamma_k$.
\begin{theorem}
For $\omega\in \Gamma_k$ and $S$ satisfying \eqref{rk1}, the corresponding  KP tau-function 
satisfies
$$
0=\tau_{\omega}^S\left(\sum_{i=1}^{k-1}\mshift{\lambda_i}\right)
$$
for any values of $\lambda_i$ ($1\leq i\leq k-1$).
\end{theorem}
\begin{proof}
The expression on the right is equal to the coefficient of $e_{0,1}$ in $\hat T\omega$ where $T=\prod(I+\lambda_iS)$.  By assumption, $\omega=v_1\wedge v_2\wedge\cdots$ where $v_i\in H_-$ for $i\leq k$.  However, since $(SH_-)_+$ is only one dimensional, all of the terms $(Tv_i)_+$ for $i\leq k$ lie in a $k-1$ dimensional subspace and hence their wedge product is equal to zero.
\end{proof}

\subsection{A  3-Term Alternative to the Pl\"ucker Relations}\label{sec:alggeom}

Although it is certainly well known that the KP hierarchy allows one
to characterize points in a Grassmannian, the approach of the present
paper provides a way to achieve this in the language of GCP maps.
Below, we will demonstrate such an approach using the standard
generator $\S$ (although any $(k,n)$-faithful $S$ would do), resulting in a single, parameter dependent, 3-term
Pl\"ucker relation that characterizes an arbitrary finite
Grassmannian\footnote{T. Shiota has shown us in personal
correspondence a possibly related procedure for characterizing an
arbitrary Grassmannian using a finite number of parameter-free 3-term
Pl\"ucker relations.}

Consider an arbitrary point $\omega\in\bigwedge^k\C^n$ and the
question of whether $\omega$ lies in the Grassmann cone
$\Gamma^{k,n}$.  Let $G$ be the $n\times n$, lower-triangular Toeplitz
matrix with the parameter 1's along the diagonal and $\alpha_i$
($1\leq i\leq n-1$) on the $i^{th}$ sub-diagonal
$$
G=\left(\begin{matrix}
1&0&0&0&\cdots&0\cr
\alpha_1&1&0&0&\cdots&0\cr
\alpha_2&\alpha_1&1&0&\cdots&0\cr
\vdots&\vdots&\vdots&\vdots&\ddots&\vdots\cr
\alpha_{n-1}&\alpha_{n-2}&\alpha_{n-3}&\alpha_{n-4}&\cdots&1
\end{matrix}\right)
$$
Denote by $P$ the projection 
$$
P(e_i)=\left\{\begin{matrix}e_i&i\geq -2\cr
0&i<-2\end{matrix}\right.
$$
Also, define $M$ to be the $n\times n$ matrix whose inverse has the block decomposition
$$
M^{-1}=\frac{1}{\Delta(\lambda_1,\ldots,\lambda_4)}\left(\begin{matrix}V_1&0\cr V_2&I\end{matrix}\right)
$$
with 
$$
V_1=\left(\begin{matrix}1&-1&1&-1\cr
-\lambda_1&\lambda_2&-\lambda_3&\lambda_4\cr
\lambda_1^2&-\lambda_2^2&\lambda_3^2&-\lambda_4^2\cr
-\lambda_1^3&\lambda_2^3&-\lambda_3^3&\lambda_4^3
\end{matrix}\right)
\qquad
V_2=\left(\begin{matrix}
\lambda_1^4&-\lambda_2^4&\lambda_3^4&-\lambda_4^4\cr
-\lambda_1^5&\lambda_2^5&-\lambda_3^5&\lambda_4^5\cr
\vdots&\vdots&\vdots&\vdots\cr
(-\lambda_1)^{k+1}&-(-\lambda_2)^{k+1}& (-\lambda_3)^{k+1}&-(-\lambda_4)^{k+1}
\end{matrix}\right).
$$

The action of the operator made by composing these maps 
$$L=M\circ P\circ G:\C^n\to\C^{k+2}$$
can be extended to a map $\hat L$ from $\bigwedge^k\C^n\to\bigwedge^k\C^{k+2}$ by letting it act separately on each component of a wedge product.
Then we define $\omega'=\hat L\omega=\sum\hat\pi_I e_I$.  Note that the coordinates $\hat\pi_I$ are now polynomials in the $n+3$ parameters $\alpha_i$ and $\lambda_i$.
Finally, as we project onto $\bigwedge^2\C^4$ by considering only the six Pl\"ucker coordinates of the form $\hat\pi_I$ with $I=(i,j,2,3,\ldots,k-1)$ and $-2\leq i<j\leq 1$:
$$
\hat\omega=\sum_{-2\leq i<j\leq i} \hat\pi_{i,j,2,3,\ldots,k-1} e_i\wedge e_j.
$$

\begin{theorem}
The point $\omega\in\bigwedge^{k}\C^n$ lies in the Grassmann cone $\Gamma^{k,n}$ \textit{if and only if} $\hat\omega$ lies in $\Gamma^{2,4}$ for all values of the parameters.  In other words, the decomposability of $\omega$ is equivalent to 
$$
\hat\pi_{-2,-1}\hat\pi_{0,1}-\hat\pi_{-2,0}\hat\pi_{-1,1}+\hat\pi_{-2,1}\hat\pi_{-1,0}=0
$$
viewed as an equation in the ring of polynomials in the variables $\alpha_i$ and $\lambda_i$.
\end{theorem}
\begin{proof}
This is a consequence of the fact that with $S=\S$, the image of the map $\hat L_{2,4}$ satisfying the Pl\"ucker relation is equivalent to the HBDE and therefore satisfied if and only if $\omega\in\Gamma$.
   Here we consider the case that a point
$\omega\in\bigwedge$ is selected such that the only non-zero
coordinates $\pi_I$ are those with $I\in\I_{k,n}$ so that all infinite
matrices can be reduced to finite dimensional ones.  We simplify
matters by considering $\alpha_i$ rather than $t_i$ where the
relationship between the two is given by the formula $ G=\sum \alpha_i
\S^i=\exp(\sum t_i \S^i).  $ Moreover, as the duality map used
explicitly in the earlier construction is not easily implemented
algebraically, we skip that step here and instead have to deal with a
more complicated change of coordinates map (constructed from the
original using the classical formula for inverse matrices) and
coordinates which still satisfy \eqref{gr24} but are permuted.
\end{proof}

\section{Concluding Remarks}\label{sec:Conclusions}

We sought to determine what property of the shift matrix $\S$ utilized
in standard Sato theory accounts for its ability to produce
tau-functions from points in a Grassmannian.  It turns out that it is
the fact that $\dim[(\S H_-)_+]=1$.  This fact can be written as a rank
one condition \eqref{rk1} on the block decomposition of the operator.

Rank one conditions of many different types have appeared in papers on
integrable systems.  For instance, their role in finite dimensional
integrable systems can be seen in
\cite{BS,HaineIliev,myBetheAnsatz,Rothstein,RS,WilsonCM} and their role in
infinite dimensional integrable systems appears in papers such as
\cite{AdenCarl,CS,TMP,KG,sakh,WentingYishen}.

In fact, \cite{TMP} represented an attempt on our part to unify
and generalize many of these different forms into a single algebraic
construction.  The present paper fulfills the promise made there to
address the geometric implications.  As we have shown, the
significance of this condition in the form \eqref{rk1} is its
relationship to the existence of a GCP linear map which translates the
Pl\"ucker relations into difference equations for the function
$\tau_{\omega}^{S}$.

\bigskip

\noindent\textbf{Acknowledgments:}  The authors wish to thank the following colleagues for their comments and advice: Emma Previato, Annalisa Calini, Tom Ivey and Takahiro Shiota.

\end{document}